\documentclass[aip,rsi,amsmath,amssymb, reprint]{revtex4-1}
\usepackage{graphicx}
\usepackage{dcolumn}
\usepackage{bm}

\usepackage[utf8]{inputenc}
\usepackage[T1]{fontenc}
\usepackage{mathptmx}
\usepackage{etoolbox}
\usepackage{xcolor}

\usepackage{amsmath} 
\usepackage{textcomp} 
\usepackage[colorlinks=true, linkcolor=blue, urlcolor=blue, citecolor=blue]{hyperref}

\makeatletter
\def\@email#1#2{%
 \endgroup
 \patchcmd{\titleblock@produce}
  {\frontmatter@RRAPformat}
  {\frontmatter@RRAPformat{\produce@RRAP{*#1\href{mailto:#2}{#2}}}\frontmatter@RRAPformat}
  {}{}
}%
\makeatother
\begin{document}


\preprint{AIP/123-QED}

\title{The MuFusE Large-Volume Diamond Anvil Cell for Exploring Muon-Catalyzed Fusion at Higher Pressures and Temperatures}
%

\author{J.D.~Kalow}
 \email[Author to whom correspondence should be addressed: ]{jkalow@acceleron.energy}
 \affiliation{Acceleron Fusion, Inc., Cambridge, MA, 02139, USA}
 \affiliation{NK Labs, LLC, Cambridge, MA, 02139, USA}
\author{J.T.~Hinchen}%
 \affiliation{NK Labs, LLC, Cambridge, MA, 02139, USA}
\author{G.~Harris}
 \affiliation{Acceleron Fusion, Inc., Cambridge, MA, 02139, USA}
\author{E.~Koukina}
 \affiliation{Acceleron Fusion, Inc., Cambridge, MA, 02139, USA}
\author{D.M.~Harrington}
 \affiliation{Acceleron Fusion, Inc., Cambridge, MA, 02139, USA}
 \affiliation{NK Labs, LLC, Cambridge, MA, 02139, USA}
\author{P.C.~McDaniel}
 \affiliation{NK Labs, LLC, Cambridge, MA, 02139, USA}
\author{N.J.~Brennan}
 \affiliation{NK Labs, LLC, Cambridge, MA, 02139, USA}
\author{A.~Golossanov}
 \affiliation{Acceleron Fusion, Inc., Cambridge, MA, 02139, USA}
 \affiliation{Laboratory for Particle Physics, Paul Scherrer Institute, 5232 Villigen, Switzerland}
\author{I.D.~Spool}
 \affiliation{NK Labs, LLC, Cambridge, MA, 02139, USA}
\author{D.~Zajac}
 \affiliation{NK Labs, LLC, Cambridge, MA, 02139, USA}
 \affiliation{Northeastern University, Boston, MA, 02115, USA}
\author{M.~Mundt}
 \affiliation{NK Labs, LLC, Cambridge, MA, 02139, USA}
 \affiliation{Northeastern University, Boston, MA, 02115, USA}
\author{S.~Varner}
\affiliation{NK Labs, LLC, Cambridge, MA, 02139, USA}
 \affiliation{Acceleron Fusion, Inc., Cambridge, MA, 02139, USA}
\author{M.~Russell}
 \affiliation{Acceleron Fusion, Inc., Cambridge, MA, 02139, USA}
\author{S.~Bull}
 \affiliation{Acceleron Fusion, Inc., Cambridge, MA, 02139, USA}
\author{K.~McCormack}
 \affiliation{Acceleron Fusion, Inc., Cambridge, MA, 02139, USA}
\author{D.~Mayer}
 \affiliation{Acceleron Fusion, Inc., Cambridge, MA, 02139, USA}
\author{L.E.~Knaian}
 \affiliation{Acceleron Fusion, Inc., Cambridge, MA, 02139, USA}
\author{M.~Khandaker}
 \affiliation{Acceleron Fusion, Inc., Cambridge, MA, 02139, USA}
 \affiliation{NK Labs, LLC, Cambridge, MA, 02139, USA}
 \affiliation{Northeastern University, Boston, MA, 02115, USA}
\author{W.~Stadolnik}
 \affiliation{NK Labs, LLC, Cambridge, MA, 02139, USA}
 \affiliation{Northeastern University, Boston, MA, 02115, USA}
\author{W.R.~Cutler} 
 \affiliation{Acceleron Fusion, Inc., Cambridge, MA, 02139, USA}
 \affiliation{NK Labs, LLC, Cambridge, MA, 02139, USA}
 \affiliation{Northeastern University, Boston, MA, 02115, USA}
 \affiliation{University of Oxford, Oxford, OX1 2JD, United Kingdom}
\author{A.~Sampat}
 \affiliation{NK Labs, LLC, Cambridge, MA, 02139, USA}
\author{K.~Lau}
 \affiliation{NK Labs, LLC, Cambridge, MA, 02139, USA}
\author{J.~Betances}
 \affiliation{NK Labs, LLC, Cambridge, MA, 02139, USA}
 \affiliation{Massachusetts Institute of Technology, Cambridge, MA, 02139, USA}
\author{C.~Fagan}
 \affiliation{Torion USA, Inc., Henrietta, NY, 14467, USA}
\author{C.R.~Shmayda}
 \affiliation{Torion Plasma Corp., Barrie, ON, L4N 9A2, Canada}
\author{M.~Koch}
 \affiliation{Torion USA, Inc., Henrietta, NY, 14467, USA}
\author{K.~Payne}
 \affiliation{NK Labs, LLC, Cambridge, MA, 02139, USA}
\author{N.J.L.~MacFadden}
 \affiliation{NK Labs, LLC, Cambridge, MA, 02139, USA}
\author{J.~Simon}
 \affiliation{NK Labs, LLC, Cambridge, MA, 02139, USA}
\author{K.~Peterson}
 \affiliation{NK Labs, LLC, Cambridge, MA, 02139, USA}
\author{A.~Gami}
 \affiliation{NK Labs, LLC, Cambridge, MA, 02139, USA}
 \affiliation{Northeastern University, Boston, MA, 02115, USA}
\author{S.~Machavarapu}
 \affiliation{NK Labs, LLC, Cambridge, MA, 02139, USA}
 \affiliation{Northeastern University, Boston, MA, 02115, USA}
\author{A.~Tejeda}
 \affiliation{NK Labs, LLC, Cambridge, MA, 02139, USA}
 \affiliation{Massachusetts Institute of Technology, Cambridge, MA, 02139, USA} 
\author{J.~Katz}
 \affiliation{NK Labs, LLC, Cambridge, MA, 02139, USA}
 \affiliation{Massachusetts Institute of Technology, Cambridge, MA, 02139, USA} 
\author{J.A.~Allen}
 \affiliation{NK Labs, LLC, Cambridge, MA, 02139, USA}
\author{R.~Chaney}
 \affiliation{NK Labs, LLC, Cambridge, MA, 02139, USA}
\author{K.~Kem}
 \affiliation{NK Labs, LLC, Cambridge, MA, 02139, USA}
\author{I.~Kiniti}
 \affiliation{NK Labs, LLC, Cambridge, MA, 02139, USA}
\author{E.~Garcia Badaracco}
 \affiliation{Fermi National Accelerator Laboratory, Batavia, IL, 60510, USA}
 \affiliation{University of California, Berkeley, Berkeley, CA, 94720, USA}
\author{K.R.~Lynch}
 \affiliation{Fermi National Accelerator Laboratory, Batavia, IL, 60510, USA}
 \affiliation{York College, City University of New York, Jamaica, NY, 11451, USA}
\author{P.~Gandhi}
 \affiliation{Fermi National Accelerator Laboratory, Batavia, IL, 60510, USA}
\author{C.J.~Johnstone}
 \affiliation{Fermi National Accelerator Laboratory, Batavia, IL, 60510, USA}
\author{E.~Niner}
 \affiliation{Fermi National Accelerator Laboratory, Batavia, IL, 60510, USA}
\author{C.C.~Petitjean}
 \affiliation{Laboratory for Particle Physics, Paul Scherrer Institute, 5232 Villigen, Switzerland}
\author{A.~Antognini}
 \affiliation{Laboratory for Particle Physics, Paul Scherrer Institute, 5232 Villigen, Switzerland}
 \affiliation{Institute for Particle Physics and Astrophysics, ETH Zurich, 8093 Zurich, Switzerland}
\author{W.T.~Shmayda}
 \affiliation{Tritium Solutions, Inc., Henrietta, NY, 14467, USA}
\author{S.O.~Newburg}
 \affiliation{Acceleron Fusion, Inc., Cambridge, MA, 02139, USA}
 \affiliation{NK Labs, LLC, Cambridge, MA, 02139, USA}
\author{A.N.~Knaian}
 \affiliation{Acceleron Fusion, Inc., Cambridge, MA, 02139, USA}
 \affiliation{NK Labs, LLC, Cambridge, MA, 02139, USA}

\date{\today}

\begin{abstract}
  A new large-volume diamond anvil cell (DAC) has been developed for the Muon-catalyzed Fusion ($\mu$CF) Experiment (MuFusE), enabling the compression and heating of deuterium--tritium ($d$--$t$) mixtures to pressures and temperatures needed to advance $\mu$CF research. The MuFusE DAC achieves the large sample volumes necessary for high-precision fusion measurements while integrating cryogenic loading, all-metal sealing, and flexible bellows to maintain a secure environment during cell compression. Combined with remote pneumatic actuation and secondary containment, the DAC safely managed a 25~Ci tritium inventory while providing a clear optical path for \textit{in situ} measurements of sample pressure and composition via laser spectroscopy. Utilizing 5~mm diameter diamond anvils oriented in the path of a high-intensity muon beam, the apparatus achieved a stable sample volume of 19.2~mm$^3$ at liquid density, pressures up to 933~MPa and temperatures up to 400~K---benchmarks that significantly exceed previously reported limits for static $d$--$t$ targets.
\end{abstract}

\maketitle

\section{Introduction}

Muon-catalyzed fusion ($\mu$CF) is a process that occurs after a negatively charged muon ($\mu^-$) replaces an electron in one of the hydrogen isotopes---protium, deuterium ($d$), or tritium ($t$)~\cite{petitjean_progress_1992,jones_survey_1988,armour_muon_2007}. Because a muon is about 200 times more massive than an electron, the resulting muonic atom is more compact, since its Bohr radius is reduced by roughly the same factor. This contraction reduces the internuclear spacing in muonic molecules and greatly enhances quantum tunneling through the Coulomb barrier. Experimentally, $\mu$CF has been observed in solid, liquid, and gaseous deuterium--tritium ($d$--$t$) mixtures in the temperature ($T$) range~\cite{bystritskii_experimental_1981,caffrey_muon-catalyzed_1986} of 3~K to 800~K---far below the $T\gg10^7~\text{K}$ required for thermonuclear fusion~\cite{lawson_criteria_1957}.
 
While $\mu$CF processes can involve all hydrogen isotope combinations, the most efficient known fusion pathway involves the formation of the mixed isotopic molecular ion $(dt\mu)^+$. In this ion, the deuterium and tritium nuclei approach closely enough to fuse, producing an alpha
particle ($\alpha$) and a neutron ($n$):
\begin{equation}
d+t\rightarrow\alpha+n+17.6~\text{MeV},
\end{equation}
where $n$ carries away 14.1~MeV of kinetic energy while $\alpha$ gets the remaining 3.5~MeV. After each reaction, the $\mu^-$ is often released and can catalyze additional fusions before it decays (lifetime $\approx 2.2~\mu\text{s}$) or binds to an $\alpha$ particle~\cite{jackson_catalysis_1957,gerstein_1977,gerstein_Erratum_1978,vinitsky_resonant_1978}.

A single muon can catalyze more than 100 fusion reactions under favorable conditions, releasing significantly more energy from fusion than the muon’s own rest energy ($\approx 105.7~\text{MeV}$). Since this was demonstrated in the mid-1980s, $\mu$CF has attracted sustained interest as a potential route to controlled fusion energy. Unlike thermonuclear approaches, it does not require magnetic or inertial confinement of an ultra-hot plasma, as the reactions can occur at comparatively ordinary temperatures and pressures within a suitably strong vessel. However, because muons are energetically expensive to produce with currently available methods, the main challenges for $\mu$CF-based energy production remain to increase the number of fusions per muon and to develop more energy-efficient sources of negatively charged muons.

The efficiency of the $\mu$CF cycle is fundamentally determined by two parameters: the sticking fraction and the cycling rate ($\lambda_c$), which is the average rate at which fusion reactions occur per active $\mu^-$. The initial sticking fraction ($\omega_0$) is the probability that, immediately after a fusion event, the $\mu^-$ becomes bound to the resulting $\alpha$ particle to form a muonic helium ion $(\alpha\mu)^+$---a muonic $\alpha$ particle. The muon may become ``reactivated'' through collisional stripping as the $(\alpha\mu)^+$ slows in the $d$--$t$ mixture, freeing the muon to catalyze additional fusion reactions. The final effective sticking fraction ($\omega_s$) is the probability that the $\mu^-$ ultimately remains stuck to the $\alpha$-particle until it decays. A high $\lambda_c$ and a low $\omega_s$ are essential to maximize the number of actual fusion reactions that each muon can catalyze~\cite{ponomarev_muon_1990, froelich_muon_1992, eliezer_muon-catalyzed_1994, spencer_kelly_investigation_2021}.

The initial sticking fraction ($\omega_0$) is the probability that, immediately after a fusion event, the $\mu^-$ becomes bound to the resulting $\alpha$ particle to form a muonic helium ion $(\alpha\mu)^+$---a muonic $\alpha$ particle. The muon may become ``reactivated'' through collisional stripping as the $(\alpha\mu)^+$ slows in the $d$--$t$ mixture, freeing the muon to catalyze additional fusion reactions. The final effective sticking fraction ($\omega_s$) is the probability that the $\mu^-$ ultimately remains stuck to the $\alpha$-particle until it decays.

These key parameters exhibit a strong dependence on the physical state of the isotopic mixture. Although $\lambda_c$ in $d$--$t$ generally increases with temperature~\cite{petitjean_progress_1992}, the highest measured rates have been obtained in cryogenic solid $d$--$t$, largely due to its high density. Crucially, these experimental results~\cite{jones_observation_1986} indicate that not only does $\lambda_c$ increase at higher densities, but the effective sticking fraction $\omega_s$ actually decreases. This unexpected reduction in muon loss, combined with enhanced cycling rates, allowed the observation of up to $150\pm20$ fusions per muon~\cite{jones_observation_1986}. Interestingly, the measured values of $\omega_s$ at high density are typically 10--50\% lower than those predicted by standard theoretical models~\cite{jones_muon-catalysed_1986, rafelski_muon_1989,melezhik_recent_2001,kamimura_comprehensive_2023}, suggesting the presence of additional reactivation mechanisms~\cite{nagamine_contribution_2001}.

 The Muon-catalyzed Fusion Experiment (MuFusE) is designed to measure $\lambda_c$ and $\omega_s$ in $d$--$t$ mixtures at temperatures, pressures ($P$) and densities beyond the ranges explored in previous experiments. Historically, only a limited number of $\mu$CF experiments have reached $P\;{\scriptstyle\gtrsim}\,100$~MPa. For instance, the group at Los Alamos National Laboratory (LANL) achieved~\cite{caffrey_muon-catalyzed_1986} $P\approx300$~MPa at $T\leqslant800$~K, corresponding to 1.3 times the liquid hydrogen density ($\phi = 4.25 \times 10^{22}$~atoms/cm$^3$). Similarly, the group at the Joint Institute for Nuclear Research (JINR) reached~\cite{bom_experimental_2005} $P=200$~MPa and $T=800$~K, equivalent to $1.2\phi$. In the present work, equivalent liquid densities are calculated using the equations of state established by Souers and Richardson~\cite{souers_cryogenic_1979, souers_hydrogen_1986, richardson_fundamental_2014}. To access combinations of $T$ and  $P$ beyond these historical benchmarks, we have developed a  specialized DAC capable of containing, compressing, and heating $d$--$t$ mixtures {\it in situ} within a muon beam.

 A DAC is a well-established device in high-pressure physics~\cite{jayaraman_diamond_1983, dunstan_technology_1989}, selected for this work due to its ability to generate extreme pressures in a safe and controlled manner, while providing optical access through the diamond anvils. The sample is contained within the hole of a metallic {\it gasket} compressed between two diamonds. Cryogenic loading of DACs with hydrogen was pioneered by Mao and Bell in the late 1970s~\cite{mao_observations_1979}; since then, researchers have reported $P>1000$~GPa in various media~\cite{dubrovinskaia_terapascal_2016} and $P>400$~GPa in hydrogen~\cite{dias_observation_2017, loubeyre_synchrotron_2020}.

 The muon beam used for this experiment, the $\pi$E1 beamline at the CHRISP facility of the Paul Scherrer Institute (PSI), can be focused to 12~mm FWHM (full width at half maximum). This is significantly larger than the typical sample area in a DAC. Furthermore, due to range straggling, muons stop over several millimeters of sample thickness even with precise momentum control. Therefore, to maximize the use of the available beam fluence, it is desirable for the compressed sample volume to be as large as possible.

 More fundamentally, to probe the muon reactivation physics that drives the difference between $\omega_0$ and $\omega_s$, the cell dimensions must be large relative to the range of the 3.5~MeV muonic $\alpha$ particle. This particle’s range is approximately 0.4~mm at liquid hydrogen density and is inversely proportional to the density. It is therefore critical that the cell be large enough to ensure the majority of muonic $\alpha$ particles undergo reactivation events within the sample before striking the cell walls.

 Static high-pressure research involving tritium has seen little development since Los Alamos experiments in the 1950s~\cite{grilly_pressure-volume-temperature_1993}. While modern methods exist for compressing hydrogen to extreme pressures in significant volumes, the radioactivity of tritium necessitates strict safety protocols. This includes limiting stored mechanical energy, employing secondary containment, avoiding polymers in wetted areas, and using specific procedures to prevent the formation of tritiated water. The highest static tritium pressure previously reported was 310~MPa at 60~K, achieved in 1956 using a closed mercury capillary system~\cite{mills_melting_1956}.

 In this paper, we describe a specialized DAC developed to research $\mu$CF in $d$--$t$ at novel conditions. In our experimental campaign at PSI, the $d$--$t$ sample reached a peak $P=933$~MPa at $T=100$~K, and a peak $T=400$~K at $P=221$~MPa.

\begin{figure}
\includegraphics{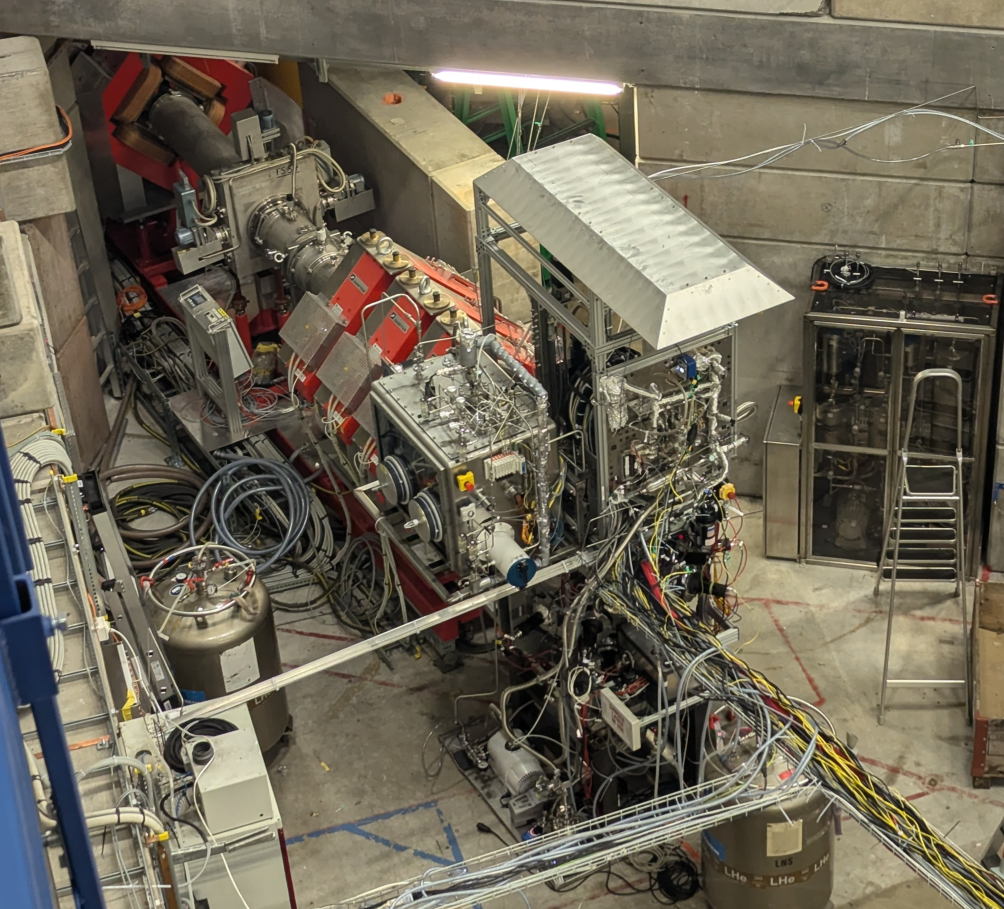}
\caption{MuFusE setup assembled in the $\pi$E1 muon beamline at the Paul Scherrer Institute in Villigen, Switzerland.}
\label{Full Experiment}
\end{figure}

\section{MuFusE Setup}

\begin{figure}
\includegraphics{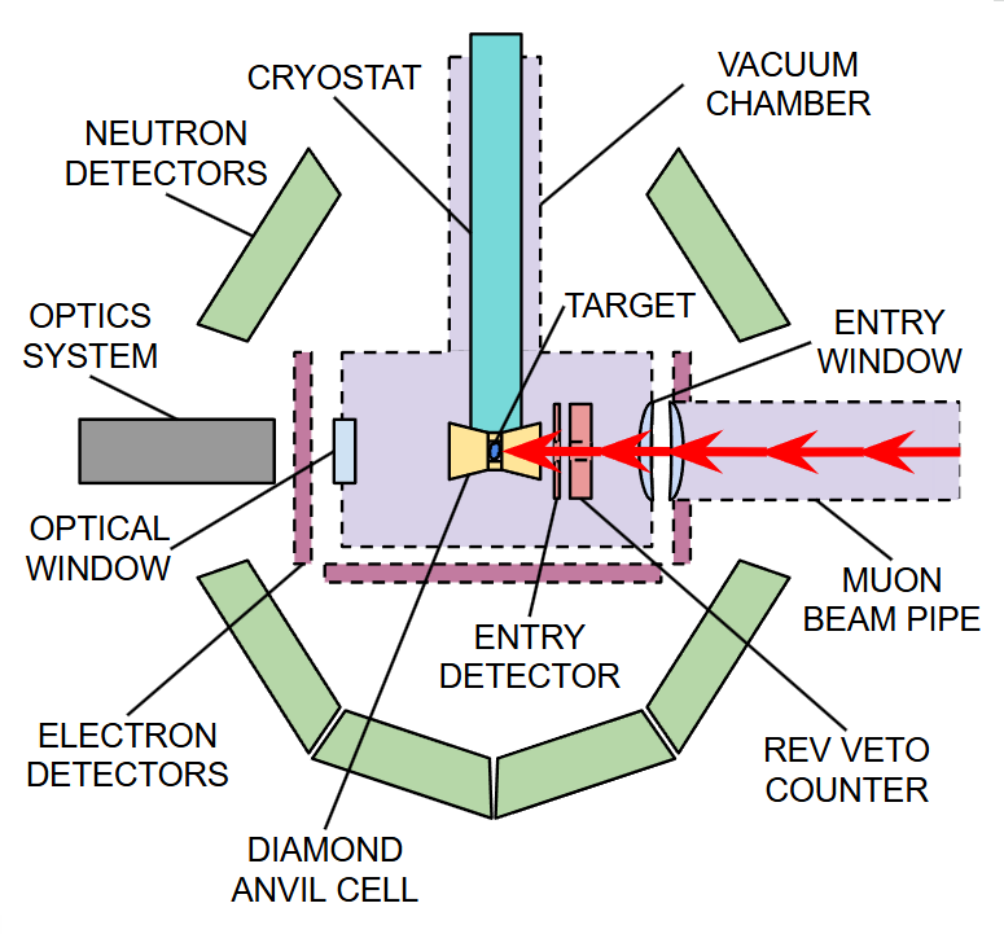}
\caption{A schematic representation of the experiment.}
\label{System outline}
\end{figure}

The DAC was at the heart of the much larger experimental apparatus---MuFusE. The primary objective of the cell was to hold a stable, compressed sample of $d$--$t$ in front of a muon beam to initiate $\mu$CF. A photograph of the MuFusE setup and its target system, placed in front of the muon beam, can be seen in Fig.~\ref{Full Experiment}. A simplified drawing showing the main components of the experiment is presented in Fig.~\ref{System outline}.

The muon beam entered the main vacuum chamber through a foil window and passed through a thin plastic scintillation counter called the entry detector, which registered each incoming muon. The muon beam then passed through the diamond anvil and into the $d$--$t$ sample. Most of the muons stopped in the cell body or in one of the diamond anvils; only a small fraction stopped in the $d$--$t$ sample itself. To simplify data analysis, the muon beam intensity was set low enough that each muon typically decayed before the next muon entered, so that muon-stopping events in each material were separated in time (to avoid pile-up). The muon beam momentum was set to 47~MeV/c, so that the location of the Bragg peak, the distance of maximum muon stopping, was focused on the $d$--$t$ sample between the diamonds.

Fast neutrons, fast electrons, and gamma rays created by $\mu$CF and other muon-induced nuclear processes passed through the DAC structure and vacuum-chamber walls and were recorded by a spherical array of scintillation counters which surround the vacuum chamber. An optical diagnostics system, which looked through the diamond anvil opposite the beam, was used to measure the pressure and composition of the gas inside the DAC in real-time. Motion stages were used to maintain precise alignment of the target, sensors, and detectors with the muon beam. A gas delivery system was designed to store, purify, and deliver deuterium--tritium mixtures to the DAC~\cite{koukina_design_2026}. A liquid-helium cryostat was used to cool the DAC to cryogenic temperatures. The main vacuum chamber surrounding the DAC provided thermal insulation, a clear path for muons, as well as secondary containment extending to the gas delivery system during tritium operations. Electric heaters were used to heat the DAC sample to elevated temperatures. The overall design of the experiment and of each subsystem will be described in more detail in other papers.

The MuFusE DAC had several design requirements. First, it needed to be capable of compressing the $d$--$t$ sample to at least $2\phi$, requiring a pressure on the order of $P\sim1$~GPa. The sample thickness needed to be at least 2~mm, and the sample diameter needed to be at least 2~mm larger than the diameter of the muon beam upon entering the cell. This is a much lower pressure, but a much higher sample volume, than most diamond anvil cells. The cell had to sustain a range of temperatures from 20 K to 500 K, while remaining hermetically sealed. The components wetted by tritium had to be constructed of materials which would not introduce ppb-level contamination through outgassing or initiate chemical reactions stimulated by beta-particles. To generate pressure in the sample, the cell components had to be movable and capable of extreme stresses at both cryogenic and elevated temperatures. Importantly, systems to measure and control the temperature, pressure, and gas composition had to be integrated with the cell and remain operable over the wide experimental range.

\section{DAC System}


In order to safely densify, heat, and maintain the target sample at extreme conditions, several important systems were implemented into the DAC. Two stainless steel gas inlet/outlet tubes were used to interface with the gas system to allow for cleaning, filling, and dumping operations. A stainless steel flexible “minichamber” contained the cryogenic liquid during loading and allowed for cell movement. With the exception of the diamond anvils, all tritium-facing components and seals had to be fully metallic or ceramic to maintain good performance and safety at extreme conditions. The diamond anvils were actuated using a high-pressure, remotely-operated, helium-filled membrane. A cell frame made from beryllium copper or maraging steel was built to house the anvil cell components and allow for the controlled movement of a piston to transmit the force from the membranes to the target. The target itself uses a deformable metal gasket sandwiched between the two diamond anvils, which traps the sample, as is common with other DACs. The anvils are mounted on matching seats, which transmit the force from the piston and cell frame. A liquid helium cryostat, used in combination with temperature sensors and heaters inserted into the seats, enabled thermal control of the sample from 10 K to 500 K.

Earlier revisions and copies of this DAC design were tested using mineral oil, argon, hydrogen, and deuterium prior to compressing cells with deuterium-tritium. Several deuterium compressions were performed in front of the muon beam and detector system at PSI for the purpose of studying $dd$ $\mu$CF. However, this work is not covered in this paper.

\subsection{DAC Target Chamber}

The target chamber size for the DAC was dictated by the size of the diamond anvils themselves as well as the size and geometry of the deformable gasket. Larger diamond anvils can quickly incur excessive costs in development, and there is a predicted inverse square relationship between culet diameter and maximum achievable pressure\cite{dunstan_technology_1989}. Some of the largest volume DACs in literature\cite{boehler_large-volume_2013} use samples smaller than $4.5*10^{-2} $mm$^3$, and culet diameters are typically\cite{obannon_contributed_2018} less than 1~mm.

\begin{figure}
\includegraphics{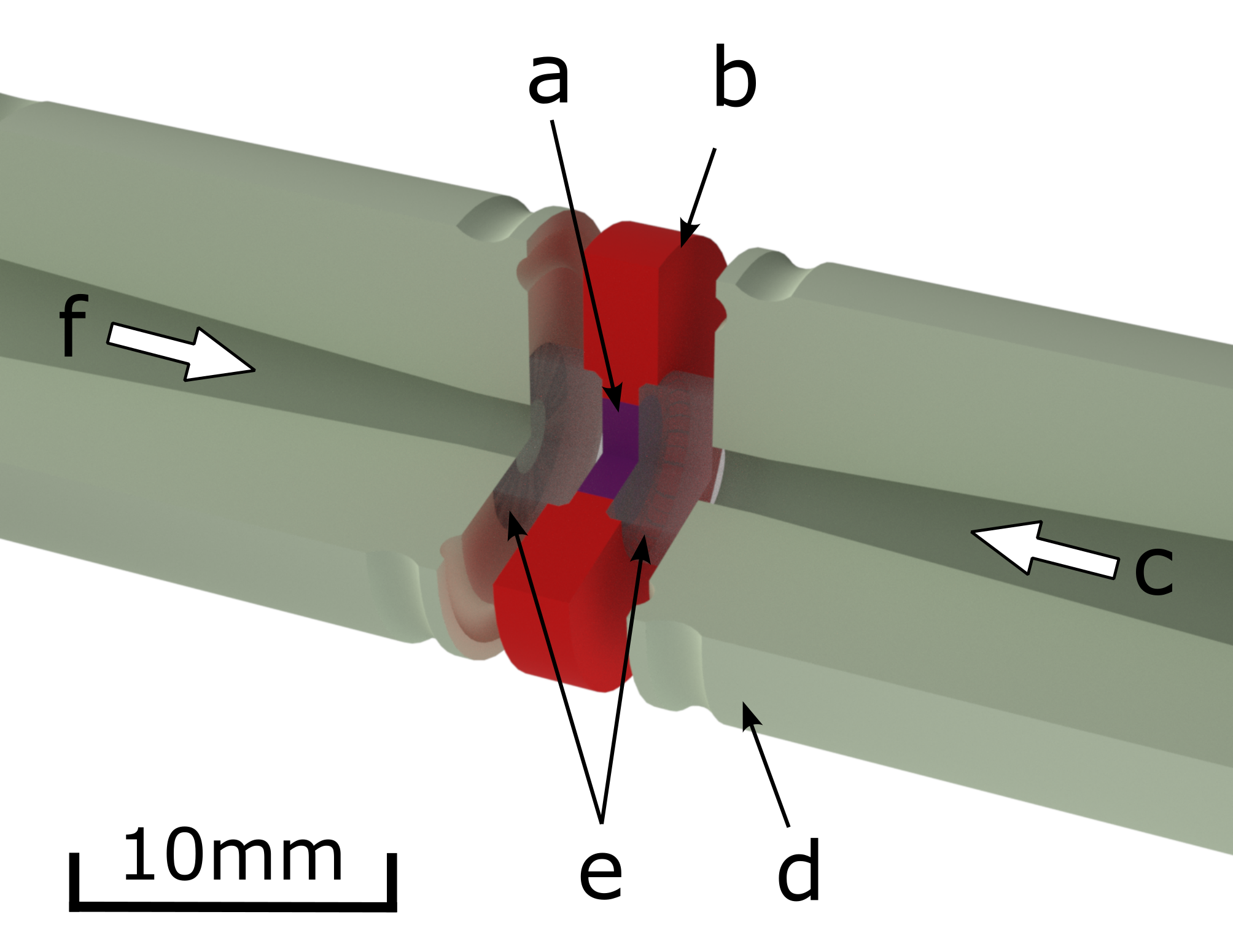}
\caption{Compressed target assembly, showing (a) compressed hydrogen target, (b) gasket, (c) muon entry window, (d) seat, (e) diamond anvils, and (f) optics entry window.}
\label{target}
\end{figure}

The target in our DAC was designed with an initial diameter ranging from 2.5~mm to 4~mm, and an initial thickness ranging from 1~mm to 2~mm, resulting in initial sample volumes ranging from 5 ~mm$^3$ to 25 mm$^3$, significantly larger than typical DACs. During compression of liquid hydrogen or $d$--$t$, this volume can be compressed by 50\% or more. A compressed target within the cell is visualized in Figure~\ref{target}.

\subsection{DAC Structure}

The anvil geometry was designed to support a load of 5~GPa across a 5~mm diameter cylindrical anvil. A small 0.2~mm chamfer was added to avoid stress concentrations on the top corner, resulting in a 4.6~mm culet. This large culet allowed for reliable compression of gaskets with holes up to 3.5~mm in diameter. While there are significant optimizations available with more complex anvil shapes (e.g. Boehler-Almax cut) for maximizing pressure, cylindrical anvils were used to achieve a very large culet size at a low cost and with minimal production time between design iterations. Each diamond anvil cost about 800 to 1200~USD, and was fabricated by either chemical vapor deposition or high-pressure high-temperature processes. The anvils were shaped with a combination of laser cutting and polishing.

The gaskets used in the tritium experiments were constructed using Aluminum~6061 and Copper~110. This was in contrast to most modern research using DACs for extreme pressure applications, which use much harder gasket materials, such as beryllium copper, stainless steel, nickel alloy, or rhenium. Our testing with aluminum and copper alloy gaskets allowed pressure generation up to 1.2~GPa with consistent and reliable operation in these large cells. Prior tests using this DAC to compress hydrogen and deuterium with harder gasket materials resulted in significantly higher pressures at high temperatures, but microscopic damage observed on the anvils has prevented safe use with tritium so far.

\begin{figure}
\includegraphics{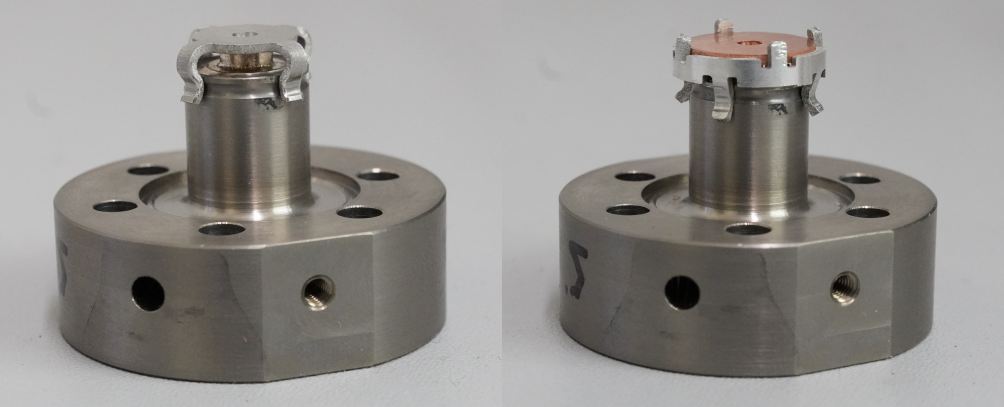}
\caption{Seats with different gasket styles.}
\label{Seats}
\end{figure}

The axis of the anvil cell was oriented parallel to the muon beam, to minimize muon scattering prior to reaching the deuterium-tritium target. The gasket needed to be placed, centered, and filled in place prior to cell compression in this horizontal orientation. A clip feature was designed to retain and center the gasket to the anvil seats. For some aluminum gaskets, this clip was integrated into the gasket itself. For other gaskets, the retention clip was a separate part constructed from aluminum. Some of these clip and gasket designs are shown mounted to their seats in Fig.\ref{Seats}.

\begin{figure}
\includegraphics{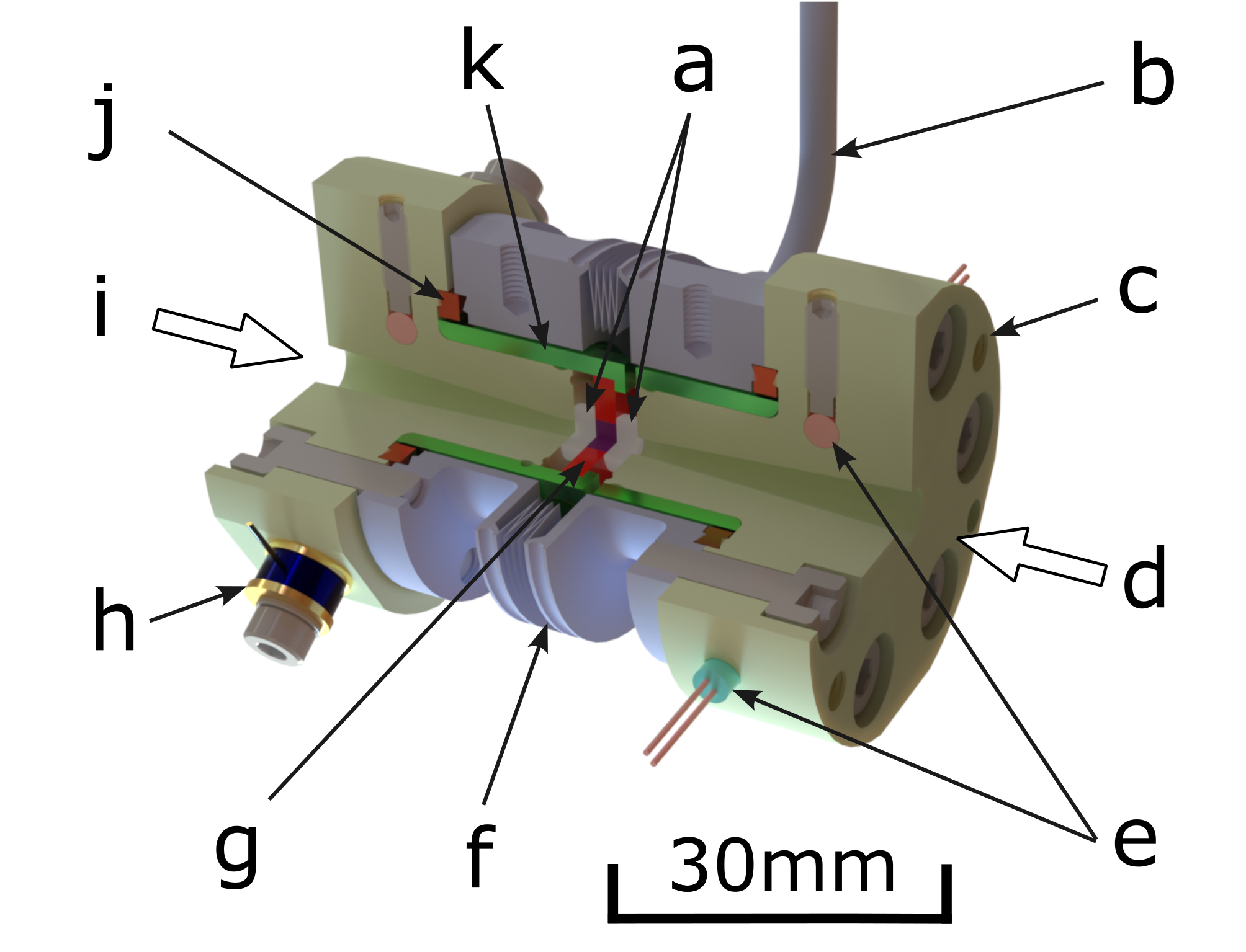}
\caption{Minichamber assembly, showing (a) diamond anvils, (b) fill tube, (c) seat, (d) muon entry window, (e) cartridge heaters, (f) bellows minichamber, (g) target, (h) temperature sensor, (i) optics entry window, (j) copper seal, and (k) gasket retainer.}
\label{minichamber}
\end{figure}

The anvils, mounting seats, and a "minichamber" were fastened together into a hermetic assembly in which liquid $d$--$t$ could be condensed. After filling the gasket hole between anvils with liquid cryogen, the DAC could be actuated, capturing and compressing the target fluid entirely within the anvils and gasket. The full minichamber assembly is shown in Fig.\ref{minichamber}.

The use of a minichamber was inspired by the work of Silvera and Wijngaarden, who performed cryogenic loading of hydrogen into DACs\cite{silvera_diamond_1985}. The use of a minichamber helped minimize the total mass of sample required to load the DAC, and cryogenic loading allowed for safe loading at low pressures, {\it in situ} in the vacuum chamber in front of the muon beam and detector apparatus. Alternative designs in literature using cryogenic loading often required the DAC to be loaded and compressed inside of a liquid helium dewar, then the cell under pressure is transferred from the dewar to a detector apparatus. The minichamber design helped greatly mitigate the risks associated with tritium - allowing for low-pressure cryogenic loading and consistent maintenance of a secondary containment from loading through to total evacuation of the cell.

Each anvil seat was constructed from either tungsten heavy alloy (MT-185) or Titanium-Zirconium-Molybdenum alloy (TZM). The primary function of the seats was to provide a strong mounting surface and conical viewing window through the diamond anvils. Both tungsten alloy and TZM allowed for a combination of high strength at high temperature, good thermal stability, and high thermal conductivity. The seats also provided holes for cartridge heaters and flats for temperature sensors.

\begin{figure}
\includegraphics{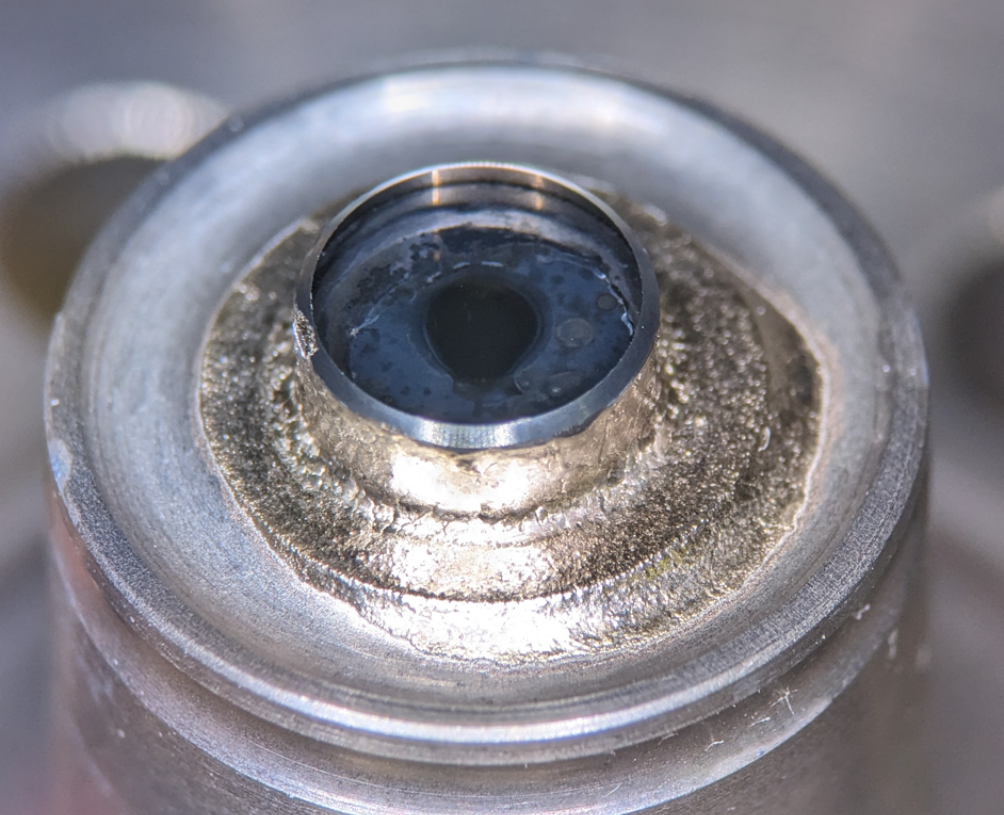}
\caption{Diamond anvil shown brazed to seat.}
\label{braze}
\end{figure}

Anvil alignment and bonding was performed using a brazing procedure developed by our group. Since the seal between the diamond anvil and the seat is part of the primary containment during filling operations, a hermetic, robust, all-metallic seal was required. This braze joint is shown in Fig.~\ref{braze}. In addition to the propensity for hydrogen gas to leak or transit through polymeric seals, the beta-particles produced by tritium decay stimulate chemical reactions which make the use of polymer seals unsuitable.

To form a strong, hermetic braze between the diamond anvils and seats, an active brazing alloy (Ticusil) was used. This brazing procedure was performed in a vacuum furnace at $10^{-4}$ mbar, using an assembly fixture to align the two parts. These braze joints were vacuum tested using a helium leak detector to a leak rate under $10^{-9}$mbar$\cdot$l$/$s, and pressure tested under ambient and cryogenic conditions with helium in the experimental chamber under high vacuum.

Previous minichamber designs by Silvera et al. used deformable indium rings to maintain a seal during cryogenic loading and cell compression. However, our cell required significantly more travel than previous hydrogen DAC designs in literature, making for an unreliable moving seal during compression. Thus, a stainless steel bellows was integrated into the minichamber to allow for hermetic sealing combined with significant movement. The bellows were welded between two machined endpieces, each featuring a conflat seal, tapped holes for the fastening of temperature sensors, and holes for stainless steel tubes to be brazed to. One tube was connected to the gas delivery system to provide both inert gas during flushing operations, and hydrogen isotope gas during filling operations. The second tube acted as a vacuum line during the bakeout and flushing process. At both ends of the minichamber, a knife edge seal and threaded bolt holes were machined in, which allowed the minichamber to be sealed to the seats using an annealed copper gasket. The copper knife edge seal, welded bellows, and brazed tube joints were all tested under vacuum using a helium leak detector to $10^{-9}$mbar$\cdot$l$/$s, and tested at high pressure up to 10~bar under cryogenic conditions with helium and hydrogen in the experimental chamber, under high vacuum.

Cell cleanout was critical to reduce impurities present in the sample during $\mu$CF measurements. Cell parts were cleaned manually and ultrasonically in baths of acetone followed by baths of isopropyl alcohol and ultrapure water. The assembly of the mini-chamber was performed inside a cleanroom hood. Prior to loading, both the cell and connecting gas lines were heated to 150$^{\circ}$C while under vacuum to bake out water and hydrocarbon impurities.

\begin{figure}
\includegraphics{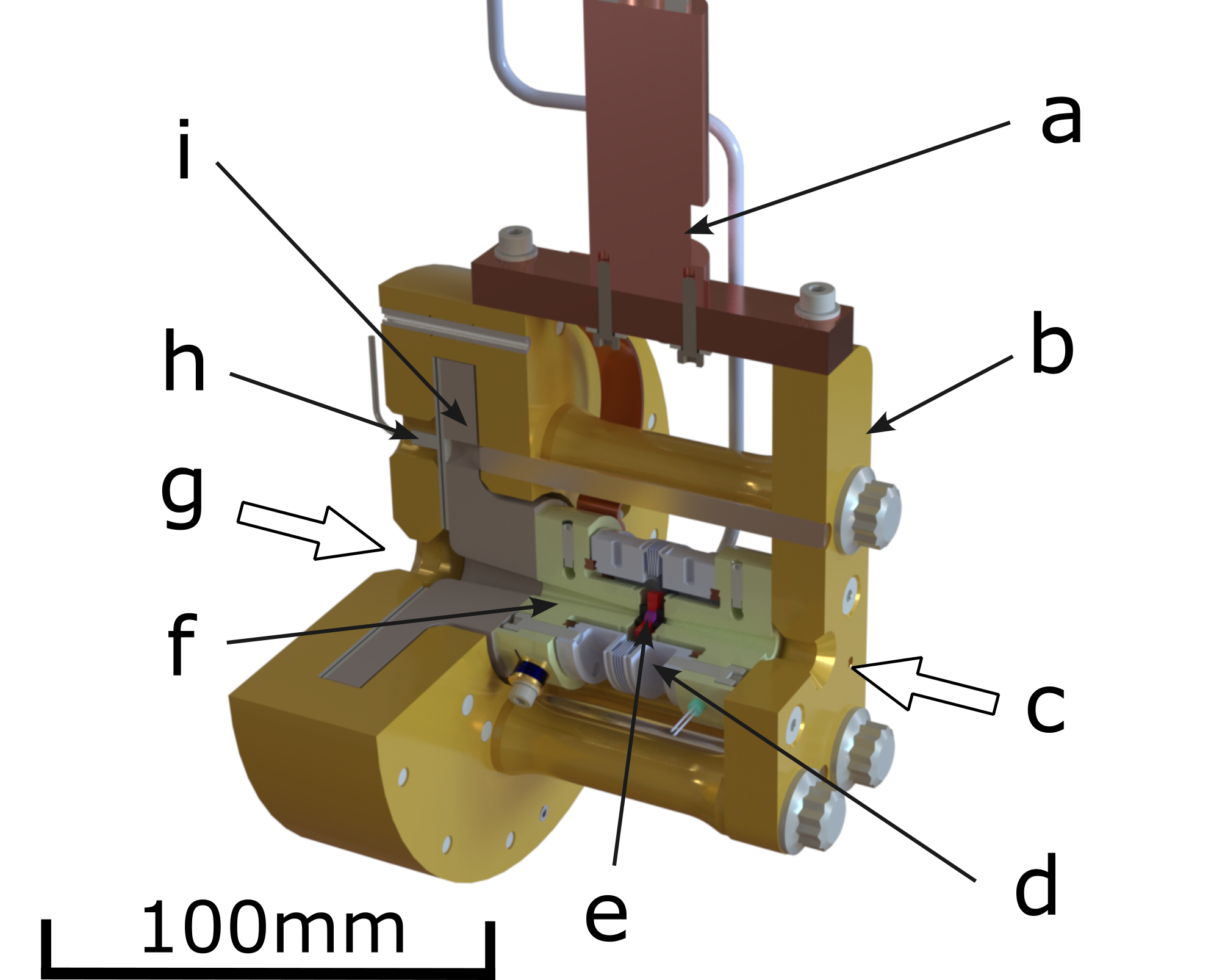}
\caption{Diamond anvil cell cross section showing (a) liquid helium cryostat, (b) cell body, (c) muon beam entry window, (d) minichamber, (e) target, (f) seat, (g) optics entry window, (h) helium membrane, and (i) piston.}
\label{DAC cross section}
\end{figure}

The main cell structure surrounding the seats and minichamber was built using components machined from copper 101, beryllium copper 172, and maraging steel C300. Copper and beryllium copper were used for their high thermal conductivity, which allowed efficient cooling of the cell to reach cryogenic temperatures. Beryllium copper and maraging steel were used because of their high strength to carry the enormous loads produced by the action of the anvil cell. While maraging steel has a much lower thermal conductivity than the copper alloys, it has a significantly higher yield strength - allowing slower, higher temperature runs. The maraging steel parts were typically coated with a nickel plating to prevent corrosion. The cell structure is detailed in Fig.~\ref{DAC cross section}.

\begin{figure}
\includegraphics{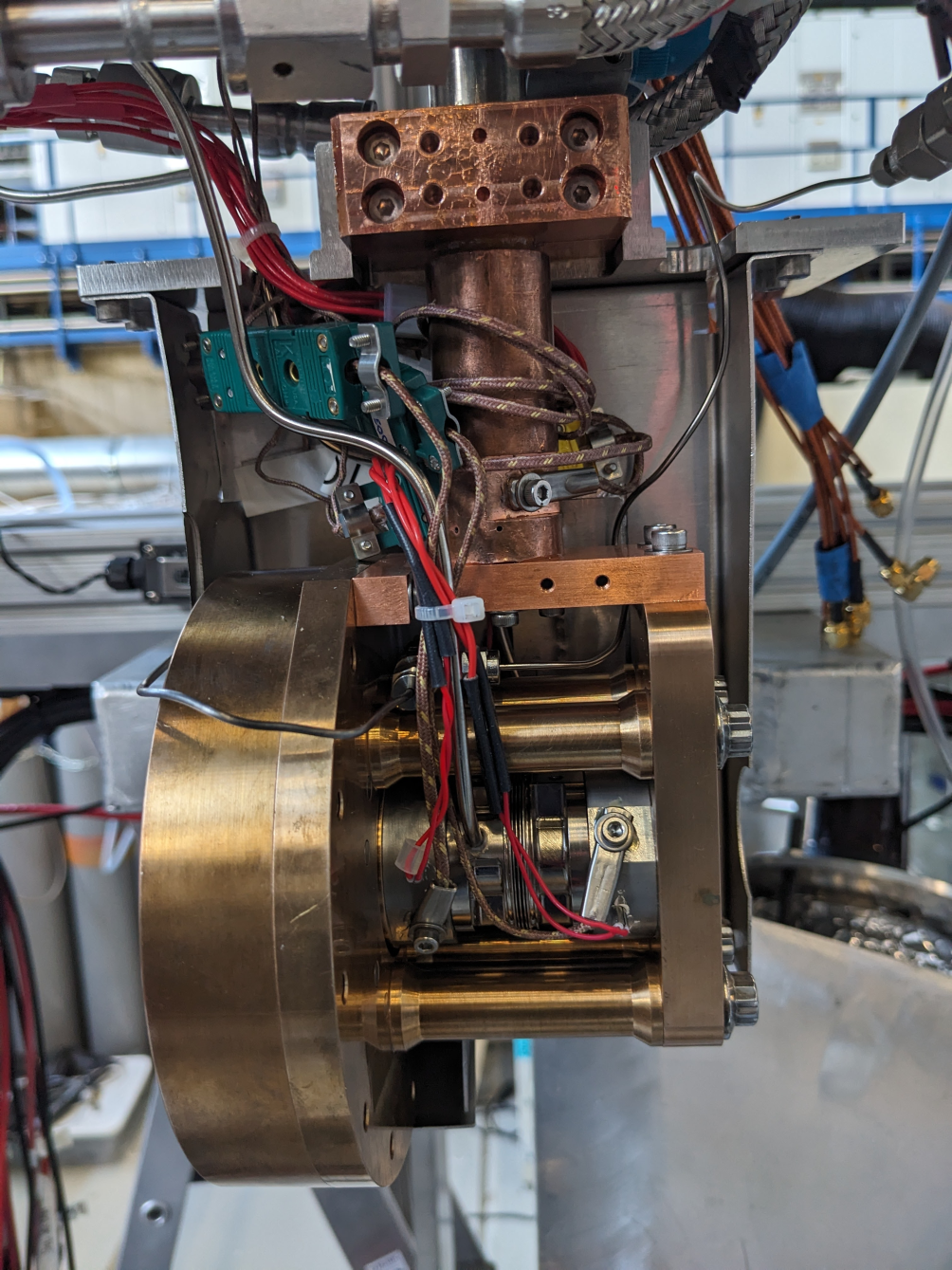}
\caption{Picture of diamond anvil cell assembled in detector system.}
\label{cell in detector}
\end{figure}

\subsection{DAC Controls}

A liquid helium cryostat from Janis/Lakeshore was used for cooling of the DAC. A thin, aluminum heat shield was fastened to the cryostat exhaust piping to block and absorb ambient thermal radiation from the surrounding vacuum chamber from reaching the cell. The cryostat tip had an integrated 75 W heater which could be controlled by a Lakeshore 335 temperature controller. The cell, fully assembled to the cryostat, can be seen in Fig.~\ref{cell in detector}.

Actuation of the DAC was achieved using high pressure helium membranes made from welded stainless steel. The pressure in the membranes exerted force on a piston, which transferred the force through the anvil seats to the diamond anvils and sample. The idea of using a membrane-driven cell to enable precise DAC control, even at cryogenic temperatures, was pioneered by Letoullec~\cite{letoullec_membrane_1988}. In this experiment, two membranes back-to-back were used to achieve the travel needed to achieve sufficient compression using aluminum gaskets. The stainless steel membranes were designed and built by Stanislav Sinogeikin at DAC Tools. While each membrane could achieve 0.5 mm of safe travel while being pressurized up to 2000~psi, the gasket deformation during runs with 2~mm thick gaskets to over 500~MPa required closer to 1~mm of travel, on top of a small spacing gap required for liquid loading. Helium pressure in the membranes was supplied from 200~bar high-purity standard gas bottles, distributed through an electronic Equilibar regulator and a 1/16’’ diameter high-pressure stainless capillary. The capillary line running from the regulator to the cell was integrated through a 3-way valve to a vacuum pump, which was used to pump ambient air out of the helium membrane prior to cryogenic operation. Multiple evacuations and helium flushes were required to clear the capillary of air such that the line and helium membrane would not clog with solidified air and become unusable during the loading operation.

Cell heating was supplemented by cartridge heaters inserted into conformal holes within the cell seats. The cartridge heaters used were supplied by Backer Hotwatt and MPI Morheat. Up to two heaters were installed in each seat.

Proper alignment of the detector assembly, DAC, and optics to the incoming muon beam was critical. The detector assembly was aligned to the muon beam using a laser level and known registration marks in the beamline facility. The DAC and optics were aligned optically with the detectors. The DAC was mounted to a 4-axis motion stage, and optics to a 3-axis motion stage. Due to the large temperature variations in the DAC and cryostat during filling and measurement, the DAC was carefully re-aligned during each muon beam run with the help of the optics system, which could be held at a known position.

\subsection{DAC Measurements and Operation}

Temperature of the DAC was measured using silicon diodes and type E thermocouples, with one temperature sensor on each seat. Due to the good thermal conductivity of the seat material, an interpolation of these two temperature sensors was found to be accurate enough for thermal control along the vapor pressure curve of the sample during the condensation and filling procedure. Additional temperature sensors on the cryostat and heat shield were used to manage the cryostat temperature.

\begin{figure}
\includegraphics{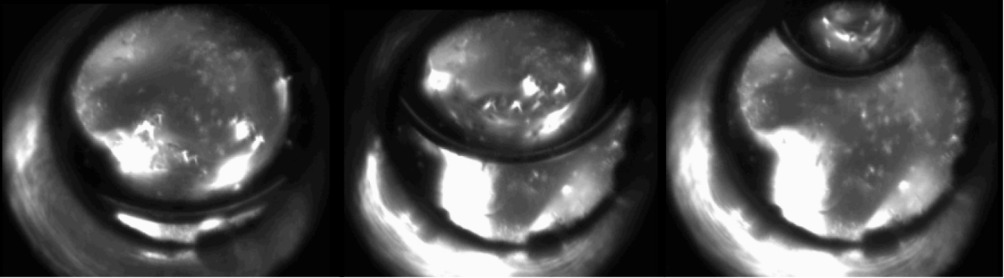}
\caption{Cell imaged during cryogenic filling. The meniscus of liquid deuterium/tritium can be seen clearly rising through a diamond anvil and 2.5mm viewing window.}
\label{Cell Fill}
\end{figure}

The cell could be optically viewed through the diamond anvil with a camera on the side opposite the muon beam entrance, attached to a remotely adjustable optical stage. An LED ring light was added opposite the camera, which provided sufficient illumination of the optical path through both anvils. The camera was used to maintain alignment between the target to the beamline following movements due to changes in cryostat temperature, and the imaging allowed for a clear view of the meniscus during the filling operation, as seen in Figure~\ref{Cell Fill}.

\begin{figure}
\includegraphics{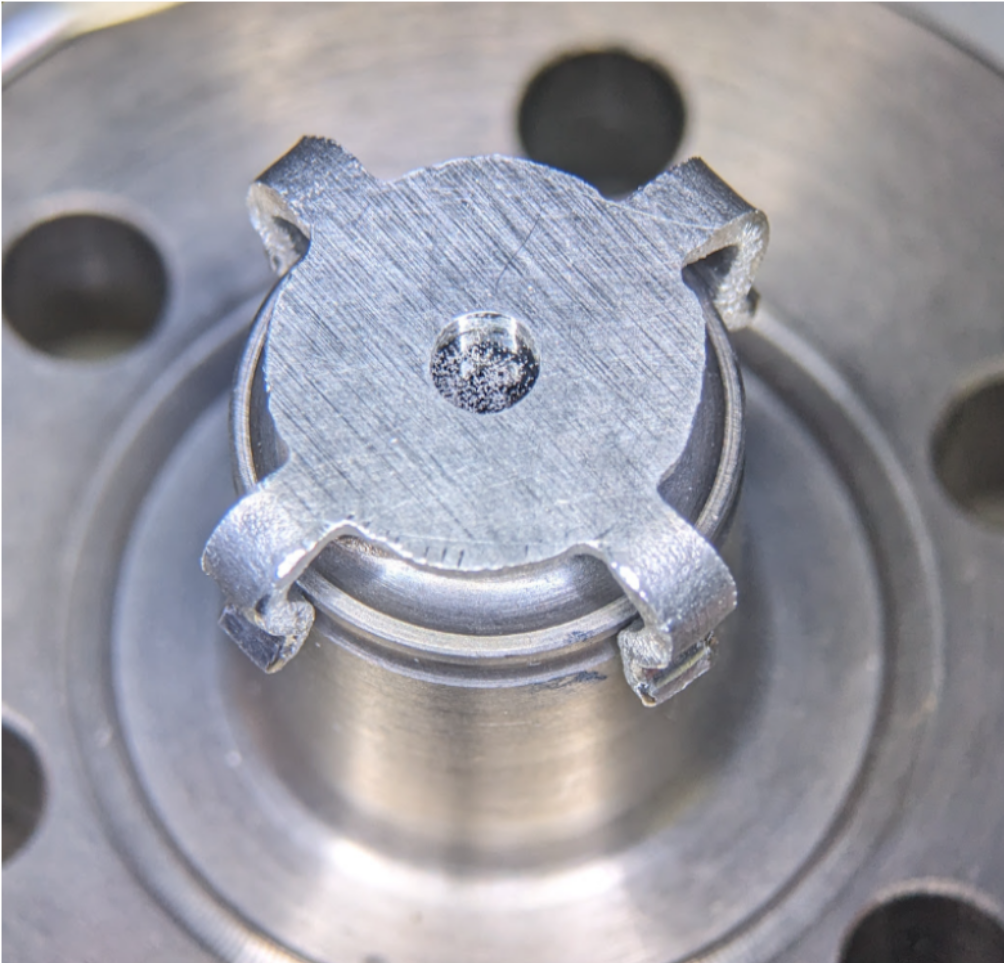}
\caption{Aluminum gasket prepared with ruby powder.}
\label{Prepped Gasket}
\end{figure}

Pressure measurement was performed using the ruby fluorescence method originally developed for DACs by Forman in 1972~\cite{forman_pressure_1972}. Small ruby particles in a size distribution from 5 to 50 microns were placed using tweezers on one anvil culet, inside of the DAC gasket. The ruby chips can be seen in Fig.~\ref{Prepped Gasket}. A 405~nm laser was used to target these ruby particles, and an Ocean Optics spectrometer was used to measure the fluorescence. The optical camera system and laser were set with the same focus point such that the camera could be used for alignment with the ruby chips. Measurements were taken continuously at each sustained interval of temperature. We adapted components of the CrimsonCalc~\cite{virant_crimsoncalc_2025} framework to perform Pseudo-Voigt peak fitting, temperature adjustment, and pressure calculation using the Ruby2020 equation to calculate cell pressure~\cite{shen_toward_2020}, and integrated them into an automated analysis pipeline to batch-process the data. Three-sigma SNR filtering was additionally applied to both fit quality (amplitude/standard error) and spectral signal (amplitude/baseline RMS). Readings outside the scope of interest (2~GPa) were visually inspected, and invalid fits were removed. Negative pressure recordings were likewise discarded. Baselines for pressure calculations were established for each run using the median of measurements collected under zero-pressure conditions.

Sample composition was measured using Raman spectroscopy. An additional spectrometer collimated with the ruby spectrometer was set up to observe the Raman shift, and the data was used to confirm and assess the filling of the DAC target chamber with the $d$--$t$ mixture.

\section{Results}

Three experimental runs with a significant balance of tritium have been performed. The first run in 2024 compressed an initial liquid volume of 4.7~mm$^3$ 60\% deuterium 40\% tritium. A maximum $P=933$~MPa was measured at $T=100$~K - equivalent to $2.4\phi$. A peak temperature of $T=400$~K was achieved at $P=221$~MPa - equivalent to $0.9\phi$.

The first $d$--$t$ test in 2025 compressed 19.2~mm$^3$ liquid volume of 50\% deuterium and 50\% tritium. A maximum $P=737$~MPa was measured at $T=125$~K - equivalent to $2\phi$. A peak $T=300$~K was achieved at $P=385$~MPa - equivalent to $1.14\phi$.

The second $d$--$t$ test in 2025 compressed 19.2~mm$^3$ liquid volume of 60\% deuterium and 40\% tritium. A maximum $P=731$~MPa was measured at $T=150$~K - equivalent to $2\phi$. A peak $T=375$~K was achieved at $P=110$~MPa - equivalent to $0.6\phi$.

\begin{figure}
\centering
\includegraphics{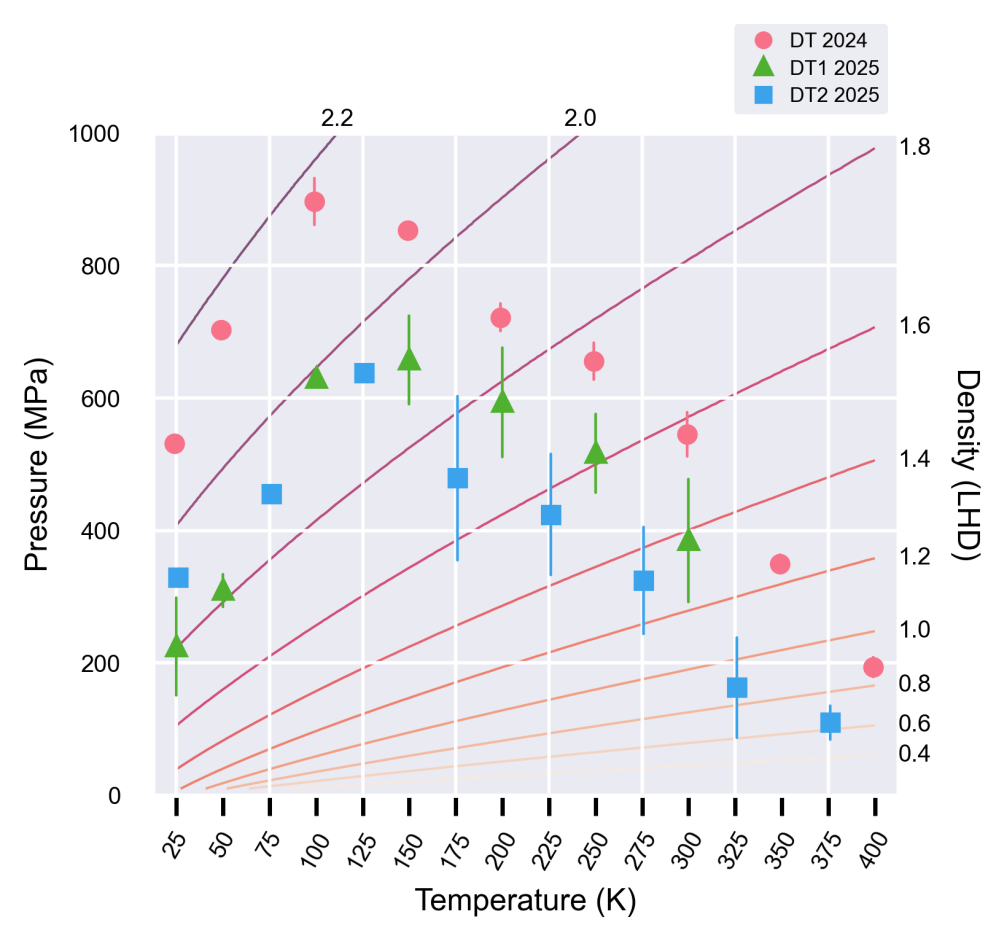}
\caption{Pressures and temperatures achieved when using a mixture of
$d$--$t$ in the DAC. Error bars represent one standard deviation. The
smooth curves are isochores---curves of constant atomic number
density---in units of liquid hydrogen density
($\phi = 4.25 \times 10^{22}$~atoms/cm$^3$), spaced $0.2\phi$ apart.
They are computed from the fundamental equation of state for
deuterium~\cite{richardson_fundamental_2014}, used here as a proxy for
the $d$--$t$ mixture.}
\label{DT2024-2025}
\end{figure}

It is notable that during these tests, as seen in Figure~\ref{DT2024-2025}, the target pressure initially rises with temperature, but then falls as the temperature ramps above 125~K. The initial increase of pressure as temperature rises is expected as the target fluid is heated in a restricted volume. However, two effects have been observed during our DAC testing which cause the pressure to later drop with increased temperature. Firstly, the yield strength of the aluminum gasket drops with increased temperature - during the heating of the compressed gasket, this results in yielding and a volume expansion of the target within the gasket. Secondly, the permeation rate of hydrogen through aluminum increases with temperature. As the sample is held at a constant temperature and high pressure during hours of fusion data collection, this permeation is observed to be significant and increasing at $T>150$~K. This permeation is observed by monitoring pressure sensors on the gas system connected to the minichamber, which safely contains the permeant during compression. While a more detailed analysis of this permeation is not yet available, it is noted that during the 2025 runs, nearly 50\% of the sample was lost through the gasket over the course of a 24 hour high-pressure run. Our DAC testing using deuterium compressed in stainless steel gaskets has shown much lower permeation rates but these materials have not been used with tritium due to safety concerns with anvils, as mentioned previously.
%
%

About 12 hours of fusion data was taken with the 60/40 $d$--$t$ mixture in 2024, 20 hours with the 50/50 mixture in 2025, and 18 hours with the 60/40 mixture in 2025. Significant $\mu$CF neutrons were observed and further analysis is ongoing.

\section{Conclusion}

We have developed a diamond anvil cell for the purposes of safely compressing a mixture of deuterium and tritium at pressures up to 933~MPa and temperatures up to 400~K. The DAC was able to compress and heat the $d$--$t$ sample while installed in a detector apparatus at a muon beam for the purposes of measuring rates of $\mu$CF at temperatures and pressures higher than previously measured.

\begin{acknowledgments}

  The information, data, or work presented herein was funded in part by the Advanced Research Projects Agency-Energy (ARPA-E), U.S.~Department of Energy, under Award Numbers DE-AR0001271 and DE-AR0001163. The views and opinions of authors expressed herein do not necessarily state or reflect those of the United States Government or any agency thereof. We wanted to thank the program management staff including S.~Hsu, A.~Diallo, L.~Chatterjee, C.~Nell, C.~Nehrkorn, M.~Handley, S.~Wurzel, H.~Jackson, R.~Wineburg, and J.~Patel.

  We wanted to thank investors and advisors, including S.~Gorbunov, W.~Lau, G.~Vlachos, C Kolster, C.~Dumas, C.~Sacca, E.~Helfgott, T.~Alberton, K.~Anson, S.~Bakalar, C.~Shapiro, F.~Nivi, B.~Hall, N.~Ravikant, M.~Sweeny, M.~Chase-Levy, R.~Surati, A.~Volpe, and N.~Heffron.

  The work has made use of facilities including the High Intensity Proton Accelerator at the Paul Scherrer Institute, the test beam facility at Fermilab, the Laboratory for Laser Energetics at the University of Rochester, Industry Lab, The Engine, and the laboratories at York College and at the Massachusetts Institute of Technology. We wanted to thank the laboratory directors and scientific advisors including K.~Kirch, A.~Amato, P.~Kammel, M.~Hildenbrandt, L.~Pedrazzi, S.~Harzmann, M.~Tisi, P.~Meyer, J.~Lykken, A.J.~Meyer, J.~McDaniel, R.~Scuzzarella, I.~Silvera, C.~Hulbert, U.~Schroder, D.~Newburg, J.~Scholvin, K.~Broderick, S.~Hudson, J. Davies, C. Forrest, M.~Kiburg, R.~Ridgeway, C.~Izzo, V.~Glebov, J.~Jacobson, and R.~Buxbaum.

We greatly appreciate all the guidance, support, and laser-welded membranes provided by S.~Sinogeikin. We gratefully acknowledge the inspiration and advice early on from leading DAC researchers R.~Boehler, I.~Silvera, M.~Lipp, and K.~De~Hantsetters.

We additionally thank all the support staff at all the institutions which have helped us. We thank all the staff at PSI for helping with the logistics, installation, integration, and repair of various parts of our experiment. Furthermore, we thank K.~Fanning and N.~Newburg at Acceleron Fusion, J.~Steve at Torion and Tritium Solutions, and A.~Van~Loon at PSI.

We appreciate all the expertise that has helped us build a novel mechanical system. J.~Bates at Bruce Diamond for help with brazing. We wanted to thank J.~Veljovic, J.~Uhl, and W.~Charuhas at Flexaseal for building us edge-welded bellows. Further thanks to E.~Shea and the team at Formex for countless machined parts, and 2fproto and DSHmold for additional machining. We thank N.~Karatzas at Drive Manufacturing, for machining our most difficult parts. For help with our diamond anvils, we wanted to thank J.~Ferguson at WD Lab-Grown Diamonds, M.~Goosen at Technodiamant, I.~Ponomarev at Adaptive Diamonds, and B.~Bailey at Element Six.

This material is based upon work supported by the Department of Energy [National Nuclear Security Administration] University of Rochester “National Inertial Confinement Fusion Program” under Award Number(s) DE-NA0004144.

This report was prepared as an account of work sponsored by an agency of the United States Government. Neither the United States Government nor any agency thereof, nor any of their employees, makes any warranty, express or implied, or assumes any legal liability or responsibility for the accuracy, completeness, or usefulness of any information, apparatus, product, or process disclosed, or represents that its use would not infringe privately owned rights. Reference herein to any specific commercial product, process, or service by trade name, trademark, manufacturer, or otherwise does not necessarily constitute or imply its endorsement, recommendation, or favoring by the United States Government or any agency thereof. The views and opinions of authors expressed herein do not necessarily state or reflect those of the United States Government or any agency thereof.

\dots.
\end{acknowledgments}

\section*{Author Declarations}

\subsection*{Conflict of Interest}

The authors have no conflicts to disclose.

\section*{Data Availability}

Data underlying the results presented in this paper are not publicly available at this time but may be obtained from the author upon reasonable request.

\bibliographystyle{aipnum4-1.bst}
\bibliography{DAC-2026}

\end{document}